\documentstyle[pra,aps,twoside,psfig,fancyheadings]{revtex}
\begin{document}
\rhead[\fancyplain{}{G. Kurizki {\it et.al.\/}: 
Optical Tachyons in \ldots}]{}
\lhead[]{\fancyplain{VII Seminar on Quantum Optics; 
Raubichi, BELARUS, May 18-20, 1998}{SQuO VII}}
\title{ Optical Tachyons in Parametric Amplifiers:
How Fast Can Quantum Information Travel?
}
\author{G. Kurizki, A. E. Kozhekin, A. G. Kofman} 
\address{Chemical Physics Department, 
Weizmann Institute of Science, Rehovot 76100 Israel}
\author{and \\ M. Blaauboer}
\address{Faculteit Natuurkunde en Sterrenkunde, Vrije Universiteit,
         De Boelelaan 1081, 1081 HV Amsterdam, The Netherlands }
\date{\today}
\maketitle
\begin{abstract}
  We show that optical tachyonic dispersion corresponding to
  superluminal (faster than-light) group velocities characterizes
  parametrically amplifying media. The turn-on of parametric
  amplification in finite media, followed by illumination by
  spectrally narrow probe wavepackets, can give rise to transient
  tachyonic wavepackets. In the stable (sub-threshold) operating
  regime of an optical phase conjugator it is possible to transmit
  probe pulses with a superluminally advanced peak, whereas conjugate
  reflection is always subluminal. In the unstable (above-threshold)
  regime, superluminal response occurs both in reflection and in
  transmission, at times preceding the onset of exponential growth due
  to the instability.  Remarkably, the quantum information transmitted
  by probe or conjugate pulses, albeit causal, is confined to times
  corresponding to superluminal velocities. These phenomena are
  explicitly analyzed for four-wave mixing, stimulated Raman
  scattering and parametric downconversion.
\end{abstract}
\draft 
\section{Introduction}

A variety of mechanisms is now known to give rise to superluminal
(faster-than-$c$) group velocities, which express the peak advancement 
of electromagnetic pulses reshaped by material media:
\begin{enumerate}
\item{{\it Near-resonant absorption}\cite{brillouin60,chu82}: 
anomalous dispersion in the linear regime of an absorbing medium 
forms the basis for this superluminal reshaping mechanism.}
\item{{\it Reduced transmission or evanescent wave formation (tunneling) in
passive dielectric structures}\cite{buettiker82,steinberg92,ranfagni93}:
this reshaping mechanism has been attributed to interference between 
multiply-reflected 
propagating pulse components in the structure\cite{japha96}.} 
\item{{\it Soliton propagation in dissipative nonlinear structures} 
\cite{picholle91}: 
superluminal group velocities can occur in such systems via nonlinear 
three-wave exchanges, as in stimulated Brillouin backscattering in the
presence of dissipation. They also occur in a nonlinear
laser amplifier\cite{icsevgi69}.}
\item{{\it Pulse propagation in transparent (non-resonant) amplifying
media}\cite{bolda96,artoni98}. Superluminal pulse reshaping
in this regime has been attributed to either the dispersion\cite{bolda96}
or the boundary reflections\cite{artoni98} of the amplifying medium.}
\item{{\it Tachyonic dispersion in inverted two-level media}:
the dispersion in such inverted {\it collective} systems is analogous to the 
tachyonic dispersion exhibited by a Klein-Gordon particle with 
imaginary mass\cite{aharonov69}. Consequently, it has been suggested
\cite{chiao96} that probe pulses in such media can exhibit 
superluminal group velocities provided they are spectrally narrow. 
Gain and loss have been assumed to be detrimental for such reshaping.
We note that Ref.~\cite{chiao96} describes an infinite medium 
and boundary effects on the reshaping have not been considered.} 
\end{enumerate}

While Ref. \cite{chiao96} suggests that tachyonic {\em
semiclassical\/} propagation is measurable in optics, it provokes
several important questions: ({\it i\/}) What other optical processes
can give rise to tachyonic behavior, and under what conditions? ({\it
ii\/}) What is the fundamental origin of tachyonic dispersion, and
does it require {\em optical coherence}? This question is prompted by
our earlier finding \cite{japha96} that coherence is the key to
superluminal group velocity of evanescent EM wavepackets (photon
tunneling) in linear dielectric media
\cite{buettiker82,steinberg92,ranfagni93}. ({\it iii\/}) Most
intriguingly, is it possible to generate {\em tachyon quanta\/} and
thereby transmit information without causality violation? The subtlety
of this question is underscored by the following consideration: If a
{\em single-photon input\/} could be transformed by a medium of length
$L$ into a {\em tachyonic quantum\/}, whose detection probability is
localized at times shorter than $L/c$, then such detection would
provide information superluminally (!).

Our search for answers to these questions has resulted in the present
theory \cite{blaauboer}, which reveals the conditions for the
observability of EM wavepackets with tachyonic features in {\em any\/}
parametrically amplifying (PA) medium. These conditions are explicitly
analyzed for processes such as stimulated Raman scattering (SRS),
parametric downconversion (PDC) and optical phase conjugation
(OPC). Two remarkable findings emerge from the present analysis, which
employs a new measure of wavepacket reshaping, namely, its
time-dependent quantum information (change in Shannon entropy): (1)
Tachyonic wavepackets transmit much less quantum information per
photon than the probe wavepackets, and their temporal resolution is
much worse than $L/c$, in accordance with causality. However, their
{\em information content is confined to superluminally short times\/},
well below $L/c$. This finding may lead to the application of
tachyonic propagation in communications. (2) Unlike evanescent EM
wavepackets in linear dielectric media, tachyonic wavepackets are {\em
insensitive to decoherence\/}.

\section{Tachyonic Dispersion in Infinite PA Media}

Our analysis will disclose the role of spatial and temporal boundary
conditions on tachyonic propagation in PA media. To this end, let us
first review the properties of an {\em infinite\/}, one dimensional
(1D) PA medium: 

\noindent
(a) The simplest model consists of two
parametrically-coupled fields at different (nondegenerate) frequency
bands, denoted by mode operators $\{a_{\vec{k}}\}$ and
$\{b_{\vec{k}}\}$ . If their coupling is effected by a classical pump
field at frequency $\omega_0$, then their interaction Hamiltonian {\em
is energy nonconserving\/}, 
$$H_{\text{int}}= - i \hbar \sum_{k\,k'} ( g_{k\,k'} a_k b_{k'} e^{i
\omega_0 t} - {\mbox h.c.} ).$$ 
It describes two-photon creation or
annihilation in correlated (``signal'' and ``idler'') modes, at the
expense of the classical ({\em undepleted\/}) pump field, whose
amplitude determines the coupling $g_{k\,k'}$. When the following
conditions hold: ({\it i\/}) the $k$-dependence (dispersion) of
$g_{k\,k'}$ is negligible, $g_{kk'}\rightarrow g$; ({\it ii\/})
phase matching prevails, and ({\it iii\/}) the uncoupled frequencies
are $\omega_{a(b) \, k} \simeq \omega_0 \pm c(k-k_0)$, we can
diagonalize this hamiltonian in the interaction picture (upon
transforming away the $\omega_0$- and $k_0$-dependence) by the
Bogoliubov transformation (BT)
\begin{eqnarray}
\label{equ2}
&D&_k = u_k a_k + v_kb^+_{-k},\quad[ D_k, H] = \hbar \omega_k D_k,\\
\label{equ3}
&u&_k=\sinh{s_k},\quad v_k=e^{i \phi} \cosh{s_k},\quad
\tanh^2{s_k}=|g|^2/|\omega_k+ck|^2. 
\end{eqnarray}
Whereas the significance of $s_k$ as the squeezing parameter is well
known \cite{barnett}, much less attention has been paid \cite{wang} to
the dispersion relation obtainable from eqs.(\ref{equ2}-\ref{equ3})
(Fig.~1)
\begin{equation}
\label{equ4}
\omega_k=\pm \sqrt{(ck)^2-|g|^2}
\end{equation}
which is {\em tachyonic\/}, i.e., isomorphic to that of a Klein-Gordon
particle with imaginary mass \cite{chiao96,kirzhnits}, and yields
superluminal group velocities $| d \omega_k / d k| \ge c$ for
spectrally narrow photonic wavepackets. 

\begin{figure}
\centerline{\psfig{file=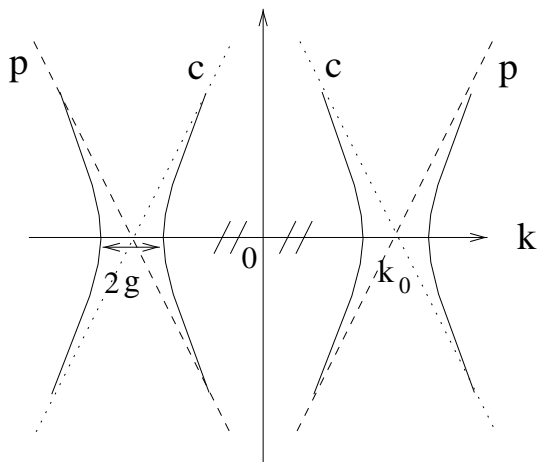,width=7cm}}
\caption{Dispersion relation (solid line) in a OPC.  The dashed
(dotted) lines correspond to the dispersion relation for the probe (p)
and conjugate (c) waves in vacuum and $k_{0} \equiv \omega_{0}/c$.}
\end{figure}

\noindent
(b) This behavior should be contrasted with that obtained when we
replace the parametric-coupling by the energy-conserving
$$H_{\text{int}} = i \hbar g \sum_k(a_k b_k^+ - \mbox{h.c.}).$$ 
This model pertains to light scattering in passive, non-dissipative
media, e.g., periodic dielectric structures with two-band
dispersion. The diagonalizing BT in this case describes passive light
reflection 
$$D_k=u_ka_k+v_k b_{-k} \,  \text{ with } \, u_k=\sin{\theta_k}, \;
v_k=e^{i \phi} \cos{\theta_k} \, \text{ and } \,
\tan^2{\theta_k}=|g|^2/|\omega_k+ck|^2.$$  
This BT yields the polaritonic dispersion
\begin{equation}
\label{equ8}
\omega_k = \pm \sqrt{(ck)^2+|g|^2}
\end{equation}
which is analogous to that of Klein-Gordon particles with real masses
(``bradyons'') and allows only subluminal group velocities $|d \omega
/ d k| \leq c$.

A more general model of infinite PA media accounts for both energy
conserving and nonconserving exchanges, and therefore pertains to a
wider range of optical parametric processes
\begin{eqnarray}
\label{equ9}
H &=& \sum_{j',j=\pm, k>0} \{ \hbar c k ( a_{jk}^+a_{jk} + \frac{1}{2}
) + \hbar \Omega_k (b_{jk}^+ b_{jk} + \frac{1}{2}) \\
\mbox{} &+& i \hbar g_k [(a_{jk}^+ b_{jk} - \mbox{h.c.})+(a_{jk}
b_{j'k} - \mbox{h.c.} ) ] \} \nonumber
\end{eqnarray}

When this 4-mode coupling Hamiltonian is diagonalized by means of a
BT, it yields a 4-branch dispersion, with two polaritonic and two
tachyonic branches. In the nondispersive limit $\Omega_k \rightarrow
\Omega_0$, this model pertains to SRS with Stokes and anti-Stokes
scattering corresponding to the energy-conserving and non-conserving
terms in eq.(\ref{equ9}). The distance between the Stokes and
anti-Stokes pairs of branches is determined by $\Omega_0$, whereas
each pair is split by $g$ (Fig.~2). Strong non-degeneracy of the two
pairs of branches $\Omega_0>g$ is required to separate between
tachyonic and bradyonic (polaritonic) types of dispersion (described
by eqs.(\ref{equ4}) and (\ref{equ8})). Otherwise, {\em hybrid\/}
(mixed tachyon-bradyon) branches can arise. When strong non-degeneracy
exists, the tachyonic branches in Fig.~2 are rotated by 45${}^\circ$
relative to those corresponding to eq.(\ref{equ4}) and are {\em
isomorphic\/} to their tachyonic counterparts in the inverted
two-level model \cite{chiao96}.

\begin{figure}
\centerline{\psfig{file=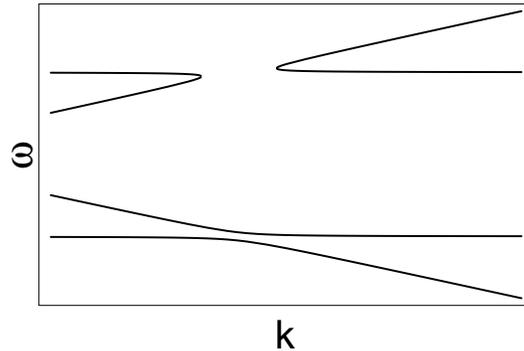,width=7cm}}
\caption{SRS dispersion obtained from a sum of the energy-conserving
  and nonconserving two-mode hamiltonians discussed in text. Upper
  curves -- Stokes dispersion (tachyonic), lower curves -- anti-Stokes
  dispersion (bradyonic or tardyonic).}
\end{figure}

The observability of the above tachyonic dispersion poses not only the
unrealistic requirement that the PA medium be infinite, but also that
we detect the field ``dressed'' by the medium, obeying
eqs. (\ref{equ2}-\ref{equ4}) (or their equivalent for
eq. (\ref{equ9})), rather than the ``bare'' field (unperturbed by
$H_{int}$). If the PA coupling is turned on at $t=0$ throughout a very
long medium, then a probe wavepacket in a multimode
(Gaussian-distributed) state will be reshaped at $t>0$ in the
dressed-field basis to exhibit superluminal peak advancement, without
any gross distortion. However, it will be accompanied by broadband
excitation of quanta arising from vacuum fluctuations within the
phase-matched spectral band, with relative weights $\langle D^+_k D_k
\rangle = \cosh^2 s_k$. In the bare-field basis, the reshaping of the
probe wavepacket at $t>0$ will be more complicated, and eventually
result in its exponential growth.

\section{Quantum Description of finite OPC}

In the rest of the paper, the above reshaping and dispersion will be
compared with those of probe photons injected into {\em finite\/} PA
media of length $L$. We shall adopt a one-dimensional description of
propagation, consistently with the standard paraxial
approximation. Since we wish to explore time-resolved information
transfer, the first generic regime to be addressed is that of {\em
abruptly switched-on pump fields\/} and short (transient) wavepacket
transmission times $L/c \lesssim t \ll \tau_{\text{rel}}$, where
$\tau_{\text{rel}}$ is the medium relaxation (or dissipation) time.
Superluminal features in this regime require the simultaneous pumping
of the entire medium, otherwise pump retardation dictates the probe
propagation velocity. Simultaneous pumping of the medium is possible
in OPC, where pump pulses can be perpendicular to the medium axis $z$
(Fig.~3-inset).

\begin{figure}
{\centerline{
\begin{tabular}{cc}
\psfig{file=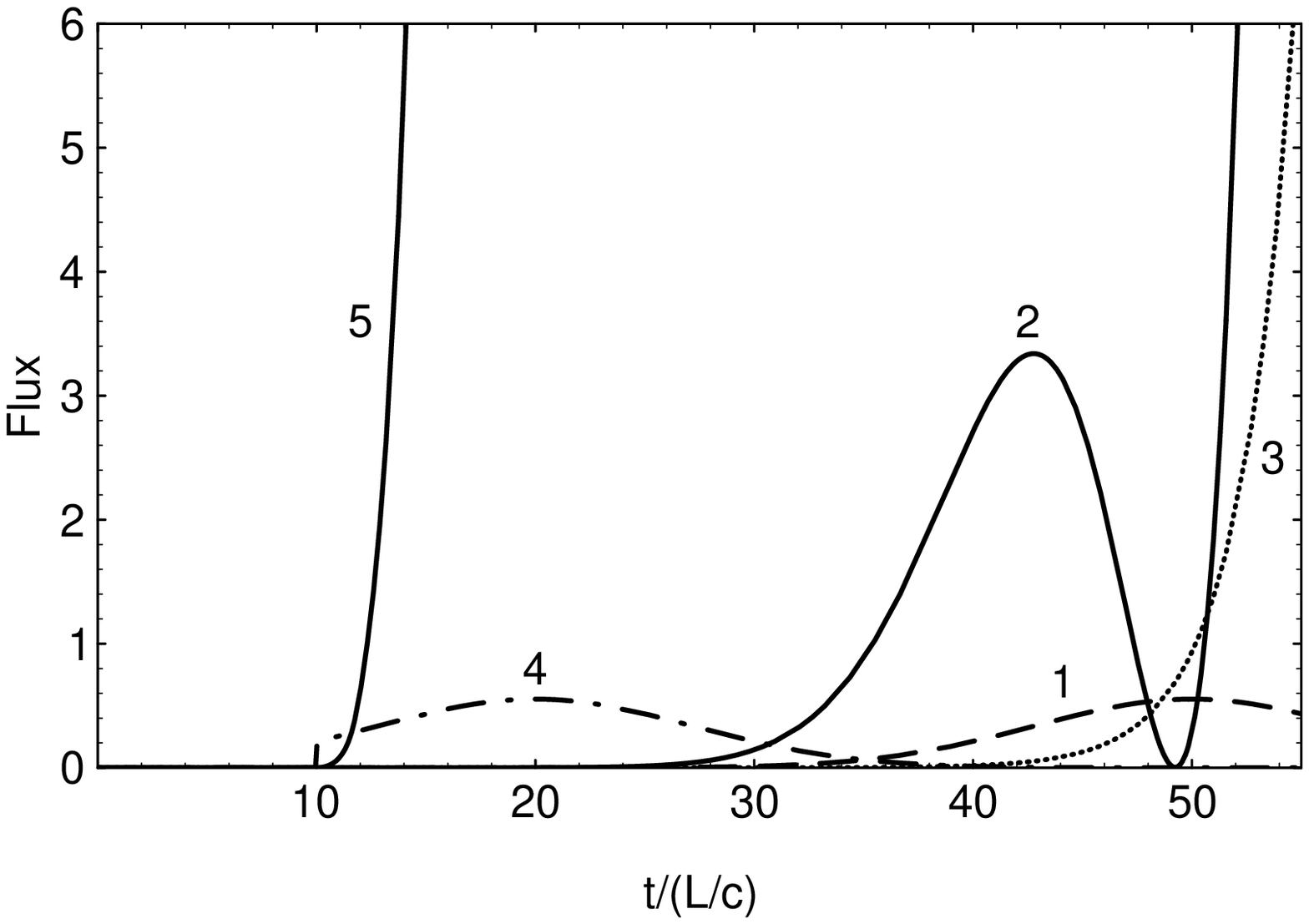,width=9cm} &
\psfig{file=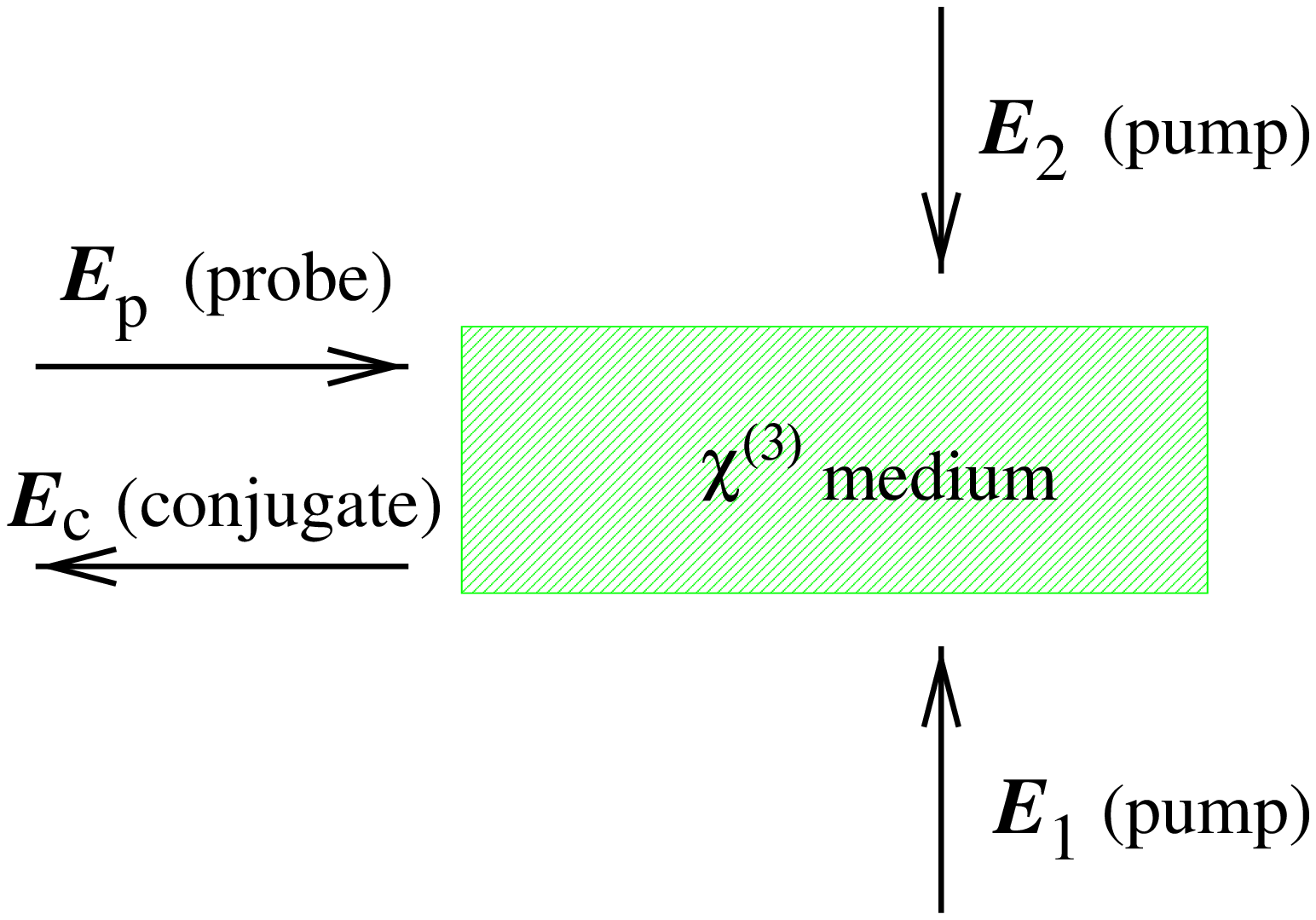,width=7cm}
\end{tabular}
}}
\caption{Conjugate (reflected) {\em wavepacket reshaping\/} in OPC:
  dependence of flux $\langle n_c \rangle$ (in units $c/L$) at $z=0$
  on time (for $|g|L=1.7$, probe width (FWHM for Gaussian amplitude)
  $\Delta_t=24L/c$, probe detuning from PA resonance $\delta=0.28c/L$,
  $t_p=50L/c$, $\langle N_p \rangle =10^{10}$); 1 -- Gaussian probe
  transmitted in empty space; 2 -- corresponding conjugate $\langle
  n_c^{(resp)} \rangle$: observe the superluminally advanced peak at
  $t \simeq 42 L/c$, followed by exponential growth (instability) at
  $t \protect\gtrsim 50 L/c$; 3 -- exponentially-growing ``noise''
  $\langle n_c^{(sp)} \rangle $; 4 -- abruptly switched-on probe
  (onset at $10L/c$, delay $t_p=20L/c$), 5 -- exponentially-growing
  conjugate corresponding to the latter.}
\end{figure}

The {\em quantized\/} analysis of OPC following the abrupt turn-on of
the pump throughout the medium has been undertaken by us for the first
time. It amounts to solving the coupled-wave equations for the slowly
varying probe ($p$) and conjugate ($c$) field operators
$\hat{E}_{p(c)}^{(\pm)}(z,t)$,
\begin{eqnarray}
\frac{\partial \hat{E}_p^{(-)}}{\partial z}+\frac{1}{c}
\frac{\partial \hat{E}_p^{(-)}}{\partial t}=ig^* \hat{E}_c^{(+)}, \nonumber\\
\frac{\partial \hat{E}_c^{(+)}}{\partial z}-\frac{1}{c}
\frac{\partial \hat{E}_c^{(+)}}{\partial t}=ig \hat{E}_p^{(-)}.
\label{8}\end{eqnarray}
Here the coupling constant $g$ is proportional to the pump intensity,
which is assumed to be turned on at $t=0$. The non-standard source
terms, i.e., $\hat{E}_p^{(-)} (z,t=0)$ and $\hat{E}_c^{(+)} (z,t=0)$,
account for the initial ($t=0$) vacuum fluctuations of the probe and
conjugate fields, respectively, at any point $z$ of the medium ($0\leq
z\leq L$). Additional sources of vacuum fluctuations are the input
fields $\hat{E}_p (z=0,t)$ and $\hat{E}_c (z=L,t)$ at the forward and
rear faces of the medium, respectively, which are taken to be in the
vacuum state. The nonvanishing correlation functions of the source
terms can be concisely expressed as
\begin{eqnarray}
\langle \hat{E}_{p(c)}^{(+)} (z,t) \hat{E}_{p(c)}^{(-)} (z',t')
\rangle_{\text{vac}}= A\delta(z-z'\mp c(t-t')),
\label{9}
\end{eqnarray}
where the upper (lower) sign refers to the probe (conjugate) field,
and $A=\hbar\omega_0 k_0^2\Omega/2\pi$, $\Omega$ being the solid angle
in which the output is detected.

We have solved Eqs.~(\ref{8}), taking account of Eq.~(\ref{9}), to
allow for spontaneous parametric generation of both probe and
conjugate fields from vacuum fluctuations, along with response to an
injected probe field $\epsilon_pf_p(t)$. Specifically $|f_p(t)|$ will
be a bell-shaped function with the maximum at $t_0$ satisfying
$|f_p(t)| \leq |f_p(t_0)|=1$. The average conjugate-reflected flux
(photons/sec) through the cross-section ${\cal S}$ and solid angle
$\Omega$ at $z=0$ is the sum of the two above components, $\langle n_c
\rangle = \langle n_c^{(sp)} \rangle + \langle n_c^{(resp)} \rangle$,
\begin{mathletters}
\label{10}
\begin{eqnarray}
\langle n_c^{(sp)}(z=0,t) \rangle = \kappa^2 \alpha\frac{c}{L}\left[
\int_0^1G_p^2(\zeta,\tau)d\zeta+\int_0^\tau
G_p^2(0,\tau')d\tau'\right], \label{10a} \\ \langle
n_c^{(resp)}(z=0,t) \rangle = \kappa^2 \alpha \langle n_p \rangle
\left|\int_0^\tau G_p(0,\tau-\tau')f_p(\tau')d\tau'\right|^2,
\label{10b}
\end{eqnarray}
\end{mathletters}
where $\kappa=|g|L$ is the dimensionless coupling strength,
$\alpha=\Omega S/\lambda_0^2$, $\lambda_0$ is the pump wavelength,
$\zeta=z/L,\ \tau=ct/L$, and $\langle n_p \rangle$ is the maximum
probe flux, $\langle n_p \rangle =\langle N_p \rangle /
\int_{-\infty}^\infty |f_p(t)|^2dt$, $N_p$ being the number of photons
in the probe pulse which pass through ${\cal S}$ into the solid angle
$\Omega$. Solution (\ref{10}) is expressed in terms of the Green
function (App.A)
\begin{eqnarray}
G_p(\zeta,\tau) = \frac{1}{2}\sum_{j=0}^\infty\left\{\left[
\left(\frac{\tau-\tau_j^+}{\tau+\tau_j^+}\right)^j
I_{2j}\left(\kappa\sqrt{\tau-\tau_j^+}\right)\right.\right.\nonumber\\
\left.\left. - \left(\frac{\tau-\tau_j^+}{\tau+\tau_j^+}\right)^{j+1}
I_{2j+2}\left(\kappa\sqrt{\tau-\tau_j^+}\right)\right]
\theta(\tau-\tau_j^+)-(\tau_j^+\rightarrow\tau_j^-)\right\}.
\label{11}\end{eqnarray}
Here $\tau_j^+=2j+\zeta$ and $\tau_j^-=2(j+1)-\zeta$ are,
respectively, the forward- and backward-propagation retardation times
after $j$ round trips, $I_{2j}()$ stands for the modified Bessel
function of $2j$-th order, and $\theta()$ for the Heaviside step
function. 

The first term in $\langle n_c^{(sp)}(t) \rangle$, Eq.(\ref{10a}), is
the response to the initial ($t=0$) vacuum fluctuations of the
probe. The second term therein represents the response to the probe
vacuum fluctuations (taking them as $\delta$-correlated) at the input
(forward) face $z=0$. The phase-conjugate response $\langle
n_c^{(resp)} \rangle$ (Eq.(\ref{10b})) to the incident probe photon
flux is {\em identical\/} with the classical response to the same
probe intensity and pulse shape $f_p(t)$. The Green function in
(\ref{11}) {\em ensures causality\/} by the step functions $\theta(
\tau - \tau_j^{\pm})$. These step functions imply that the $j=0$ term
of the response allows superluminal reshaping only via the forward
tail of the probe wavepacket extending throughout the medium at
$t=0$. The $j \neq 0$ terms in (\ref{11}) account for multiple
reflections at the boundaries.

Above the threshold for parametric instability, the spontaneous
parametric radiation $\langle n^{(sp)}_c(t) \rangle$ (triggered by the
probe vacuum fluctuations) grows at $t>0$ nearly
exponentially. Nevertheless, there can be quasi-tachyonic reshaping,
namely, the transformation of a Gaussian probe wavepacket into a
wavepacket $\langle n_c^{(resp)}(t) \rangle$, which resembles a
Gaussian with a superluminally advanced peak at times
$0<t<t_u$. The time $t=t_u$ marks the end of quasi-tachyonic reshaping
and the start of the {\em unstable classical\/} response, i.e., the
onset of exponential growth of $\langle n_c^{(resp)}(t) \rangle$, so
that at $t>t_u$ the output becomes {\em unrelated\/} to the
input. This growth is changed at $t \gtrsim \tau_{\text{rel}}$,
$\tau_{\text{rel}}$ being the relaxation time (see below). This sharp
separation of the two evolution stages: $0<t<t_u$, the tachyon
transient time, and the instability growth time
$t_u<t<\tau_{\text{rel}}$ is contingent on the following conditions.
We need a narrow spectral width of the incident probe, large detuning
of the probe central frequency from parametric resonance, and large
delay time $t_p$ between the pump turn-on $t=0$ and the arrival of the
incident-probe peak at $z=0$, otherwise there is fast onset of the
instability (Fig.~4).

\begin{figure}
\centerline{\psfig{file=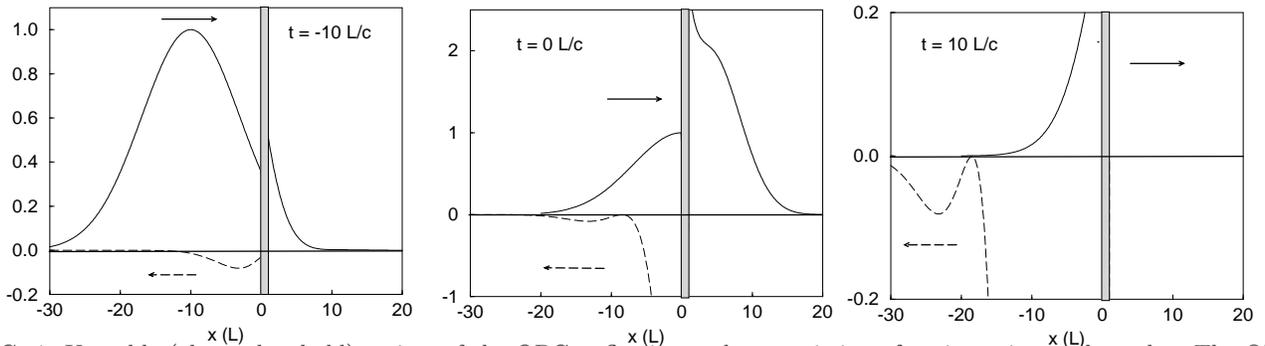,width=17cm}}
\caption{Unstable (above-threshold) regime of the OPC: reflection and
  transmission of an incoming probe pulse.  The OPC medium is located
  between $0 < x < L$ (see shaded area).  The upper (lower) half of
  each time plot shows the probe (conjugate) pulse $|{\cal
  E}_{p}(x,t)|^2$ (- $|{\cal E}_{c}(x,t)|^2$), which is moving to the
  right (left). Parameters used are $\delta_{0} = 0.31\, c/L$,
  $\Delta_{t}=19.3\, L/c$, $g L = 1.7$.  }
\label{fig:trans6}
\end{figure}

\section{Quantum Information Travel in OPC}

What is the physical significance, if any, of the quasi-tachyonic
reshaping at $0<t<t_u$? To answer this question, we have considered a
quantity that has not been used before to study wavepacket reshaping,
namely, the time-dependent quantum information function $\Delta
H_c(t)$, associated with the Shannon entropy of the detected flux
$\langle n_c(t) \rangle$. This function reads \cite{caves}
\begin{eqnarray}
\Delta H(t) &=& \langle n_c \rangle \log_2 \left( 1+ \frac{1}{ \langle
n_c \rangle } \right) \nonumber \\ \mbox{} &+& \log_2 \left( 1+
\frac{\langle n_c^{(resp)} \rangle }{ \langle n_c^{(sp)} \rangle }
\right) - \langle n_c^{(sp)} \rangle \log_2 \left( 1+ \frac{1}{
\langle n_c^{(sp)} \rangle } \right)
\label{eq:info}
\end{eqnarray}
A similar function $\Delta H_p(t)$ can be defined for the transmitted
probe $\langle n_p(t) \rangle$. When the probe wavepacket envelope
$f_p(t)$ has a sharp front, these information functions behave
causally, i.e., they grow only at times satisfying
retardation. However, when $f_p(t)$ has an {\em extended forward
tail\/}, then these functions exhibit apparent superluminal
advancement compared to the same function for a probe in free space
(Fig.~5). The reason for this is that the forward tail of the probe is
transformed into $\langle n_c^{(resp)} \rangle$ right after the pump
turn-on at $t=0$, hence {\em this information growth is due to the
nonlocal (simultaneous) pump effect throughout the medium\/}, and no
retardation is expected. The information increases until $\langle
n_c^{(sp)} \rangle$, which grows exponentially and contributes {\em
negatively\/} to $\Delta H$, becomes appreciable, compared to $\langle
n_c^{(resp)} \rangle$. This causes $\Delta H_c$ to be confined to
earlier times than the peak of $\langle n_c(t) \rangle$!

\begin{figure}
{\centerline{\psfig{file=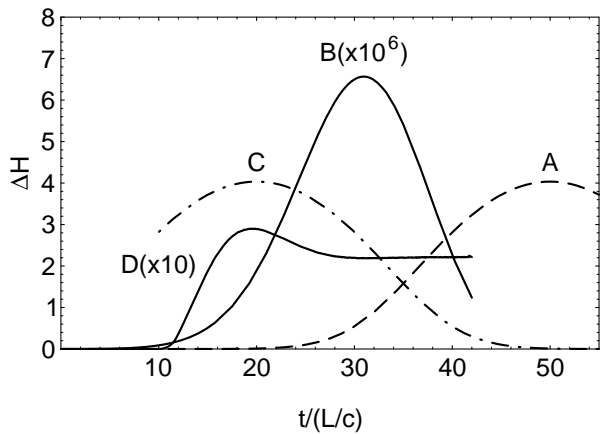,width=9cm}}}
\caption{ {\em Information transfer\/} $\Delta H_c$
  (Eq.(\ref{eq:info})), same parameters as in Fig.~3, except for
  $\langle N_p \rangle=100$; A -- for Gaussian probe in empty space
  $\Delta H_c$ follows probe shape; B -- for corresponding conjugate
  (with ``noise''), $\Delta H_c$ is peaked at an earlier time $t
  \simeq 30 L/c$, than curve 2 in Fig.~3, but is $10^6$ times smaller;
  C -- for chopped probe (onset at $10L/c$) transmitted through empty
  space; D -- for conjugate (with ``noise'') corresponding to the
  latter, note that $\Delta H_c$ is no longer advanced, but peaked at
  $t \simeq 20L/c$, just as curve C.  }
\end{figure}

Note, however, that the total amount of information transfer is
sharply reduced as compared to that of a probe in free space, and does
not exceed that of the front tail of the free-space probe, as required
by causality. Nevertheless, the temporal compression of the
information due to the nonlocal character of the response is a
potentially useful novel feature for communications: it indicates the
appropriate time intervals for modulation of the probe envelope, so as
to optimize the information transfer. As seen from Fig.~5 the Gaussian
{\em probe\/} envelope can be switched off well before it reaches its
peak, since the transferred information $\Delta H_c$ is derived only
from the forward tail! Likewise, the minimal useful interval between
successive probe peaks should exceed the temporal width of $\Delta
H_c$.

\section{Tachyonic Propagation in SRS}

In SRS, by contrast to OPC, the pump cannot be switched on
simultaneously throughout the sample rather the pump is moving along
the sample (in the $z$ direction) with a front velocity $c$.
Therefore the Stokes (signal) field in SRS or signal in PDC has
superluminal features only if the pump front sweeps the sample axis
$z$ at a speed exceeding $c$, as allowed by the experimentally
feasible ``searchlight effect''.

On the other hand, the SRS regime in which superluminal features are
displayed even for a {\em stationary\/} pump is that of {\em long
times\/}, at which transient effects due to the pump turn-on play no
role already, but relaxation effects are no longer negligible. The
slowly varying Stokes field operator in SRS for a stationary laser
pump intensity $I_L$, and relaxation (dissipation) rate
$\Gamma=1/\tau_{\text{rel}}$, is given by the known solution
\cite{raymer}
\begin{eqnarray}
\hat{E}_s^{(-)}(z,t)=\hat{E}_s^{(-)}(0,t)+ \frac{1}{2} \alpha^{1/2}
\left\{z^{1/2} \int_0^tdt'e^{-\Gamma t'} \frac{I_1[(\alpha
zt')^{1/2}]} {(I_Lt')^{1/2}} \hat{E}_s^{(-)}(0,t-t') \right. \nonumber
\\ \left.-ie^{-\Gamma t}\int_0^zdz' \hat{Q}^\dagger (z-z',0)
I_0[(\alpha z't)^{1/2}] -i \int_0^tdt' \int_0^zdz' e^{-\Gamma t'}
I_0[(\alpha z't')^{1/2}] \hat{F}^\dagger(z',t')\right\}.
\label{12}
\end{eqnarray}
Here $\alpha=g\Gamma=4\kappa^2I_L$, $\hat{F}(z,t)$ is the
$\delta$-correlated (in space and time) Langevin operator describing
the field dissipative fluctuations, and $\hat{Q}(z,0)$ is the initial
spatially $\delta$-correlated atomic inversion operator.

The plots (Fig.~6) show the output Stokes (signal) of the average
intensity $\langle n_s \rangle = \langle \hat{E}_s^{(-)}(z,t)
\hat{E}_s^{(+)}(z,t)\rangle$ obtainable from (\ref{12}) for incident
Gaussian input wavepackets with appropriately chosen width, mean
frequency and peak delay (after the pump turn-on at $t=0$), reveal a
surprising finding: quasi-tachyonic reshaping obtains for large
transit times $t\gg\tau_{\text{rel}}$ (if the response contribution to
$\langle n_s \rangle$ strongly exceeds the vacuum fluctuations
contribution, which comes from the last two terms in (\ref{12})). This
reshaping is {\em essentially similar\/} to that of the transient
($t\ll\tau_{\text{rel}}$) regime of OPC (Fig.~3). Hence, {\em
relaxation\/} (and the resulting destruction of temporal coherence)
{\em does not preclude the tachyonic reshaping\/}, as opposed to the
superluminal reshaping of optical evanescent waves, which is
essentially a coherent (interference) process \cite{japha96}. Here, by
contrast, tachyonic reshaping in the presence of dissipation
originates from the $t'$-dependence of the integrand $e^{-\Gamma
t'}I_1[(\alpha zt')^{1/2}]/\sqrt{t'}$ in (\ref{12}), which is similar
to that of a Gaussian (spectrally dispersive) function. This
conclusion can be generalized to any PA medium: for $t \gg
\tau_{\text{rel}} $, it is possible to observe tachyonic dispersion by
taking advantage of the high {\em spectral selectivity\/} of the
parametric gain coefficient, $\sim g/[1+\tau_{\text{rel}}(\omega-
\omega_0)]$, which is limited to a band of width $\sim \Gamma =
1/\tau_{\text{rel}}$. After the pump is switched on, at times $t
\gtrsim \tau_{\mbox{\scriptsize rel}}$, the Fourier-transformed
response function has the generic form $ \tilde{G}(\delta) \simeq
e^{\frac{(gL/c)}{1+(\tau_{\text{rel}} \delta)^2}} \simeq
e^{\frac{gL}{c}(1-(\tau_{\text{rel}} \delta)^2)} $, where
$\delta=\omega-\omega_0$. The nearly-Gaussian shape of the response
$\tilde{G}(\delta)$ allows for tachyonic features, when
$\tilde{G}(\delta)$ is convoluted with a Gaussian probe.

\begin{figure}
{\centerline{
\begin{tabular}{cc}
\psfig{file=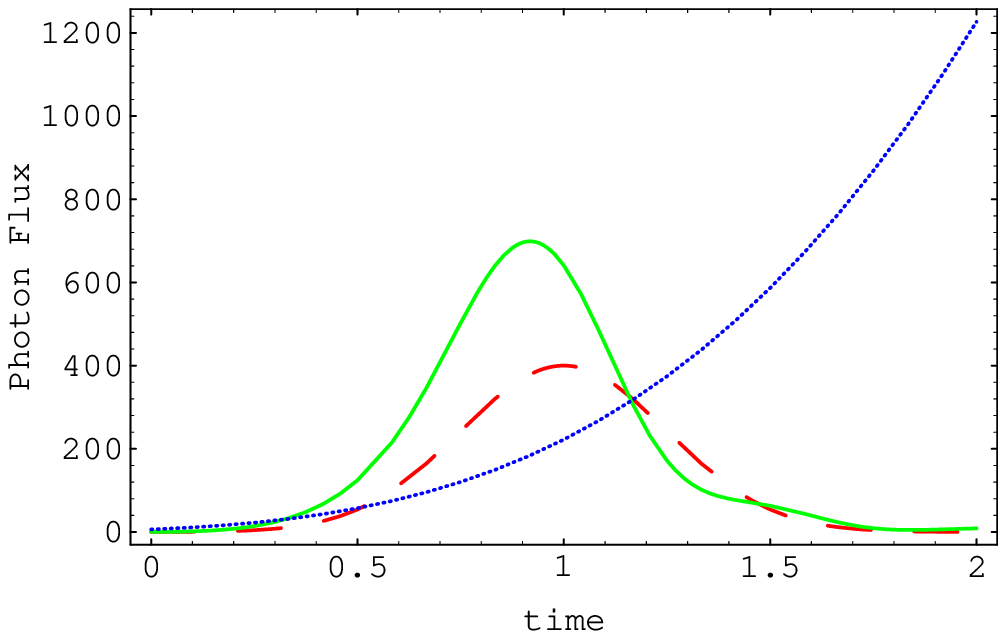,width=8.5cm} &
\psfig{file=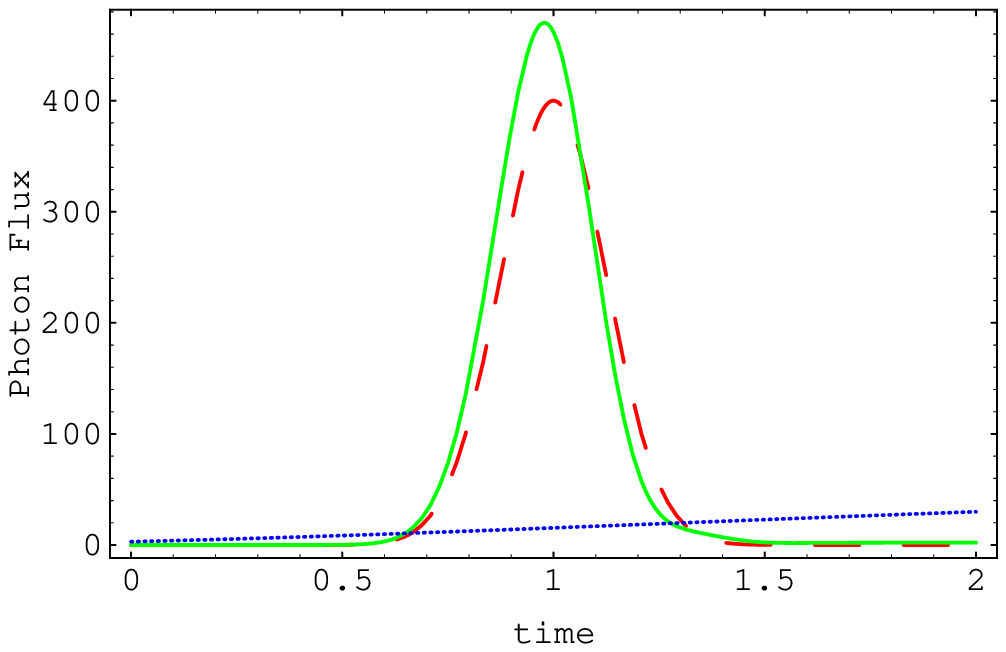,width=8.5cm} 
\end{tabular}
}}
\caption{Transmitted SRS photon flux ($\langle n_s \rangle /\Gamma$)
  as a function of time ($\Gamma t$): dotted line --
  exponentially-growing spontaneous (vacuum fluctuations) ``noise'';
  dashed line -- the probe flux in free space; solid line -- induced
  response with superluminally-advanced peak compared to dashed
  line. (a) Detuning $\delta/\Gamma=10$, probe pulse width
  $\Delta/\Gamma=0.5$, $g L = 25$; (b) detuning $\delta/\Gamma=15$,
  probe pulse width $\Delta/\Gamma=0.25$, $g L = 12$. }
\end{figure}

\section{Conclusions}

Our theory has yielded the following conclusions, which elucidate the
nature and observability of tachyon effects in quantized PAs: (a)
Tachyons are the characteristic quanta generated in {\em any
infinite\/} parametrically amplifying (PA) medium, involving processes
such as stimulated Raman scattering (SRS), parametric down conversion
(PDC) and optical phase conjugation (OPC). The basic connection
between parametrically amplifying (PA) processes and tachyonic
dispersion is that their inherent energy-nonconservation entails lack
of symmetry under time reversal, along with translational invariance
(momentum conservation or phase matching).  This signifies that the
space and time (or energy and momentum) axes in PA dispersion are
interchanged as compared to the energy-conserving, time-reversible
polaritonic dispersion, corresponding to the 90${}^\circ$ rotation of
axes when passing from a bradyonic Lorentz transformation to the
tachyonic generalized Lorentz transformation \cite{feinberg}. (b) In a PA
medium of length $L$, only probe wavepackets with {\em macroscopic\/}
photon numbers and temporal width well above $L/c$ can be reshaped
into observable (transmitted or reflected) tachyonic wavepackets
(characterized by superluminal peak advancement and little
distortion). The observability of these tachyonic wavepackets is
limited to transient times by the inherent {\em quantum instability\/}
of PA processes, i.e., the broadband amplification of vacuum
fluctuations on the one hand, and, on the other hand, by the classical
instability (exponential growth) associated with gain. (c) As a result
of competition between reshaping and instability the {\em information
content\/} of tachyonic wavepackets is confined to times well below
$L/c$, although causality is not violated. This novel feature is shown
to be advantageous for communications. (d) Tachyonic dispersion of
probe wavepackets is observable even in the dissipative, {\em
incoherent} propagation regime, by virtue of the PA gain dispersion in
the presence of dissipation.

We acknowledge useful discussions with I.\ Bialinicki-Birula and Y.\
Silberberg, and the support of Minerva and EU(TMR) grants.

\appendix
\section{The Green function derivation for OPC}

In this appendix we briefly outline the two-sided Laplace transform
(TSLT) technique introduced by Fisher {\it et al.}\cite{fisher81} for
the derivation of Green function. The TSLT is defined as $\tilde{\cal
F}_{(*)}(x,s) \equiv \int_{-\infty}^{\infty} dt\, F^{(*)}(x,t)\,
e^{-st}$. It is only valid for functions $F(x,t)$ that diminish faster
than exponentially at times $t\to -\infty$. Its advantage compared to
the the usual one-sided Laplace transform is that it also applies to
functions that do not vanish at $t<0$.  The starting point of the
analysis is to apply the TSLT to the four-wave mixing equations
(\ref{8}) with $F(x,t) = {\cal E}_{p}(x,t)$ or ${\cal
E}_{c}^{*}(x,t)$.  We then obtain the coupled equations
\begin{equation}
\left\{ \begin{array}{l} 
\frac{d}{dx}\, \tilde{\cal E}_{p}(x,s) + \frac{s}{c}\, \tilde{\cal E}_{p}(x,s)
+ i g\, \tilde{\cal E}_{c,*}(x,s) = 0 \nonumber \vspace{0.3cm}\\
\frac{d}{dx}\, \tilde{\cal E}_{c,*}(x,s) - \frac{s}{c}\, 
\tilde{\cal E}_{c,*}(x,s)
+ i g^{*}\, \tilde{\cal E}_{p}(x,s) = 0.
\end{array} \right.
\label{eq:LaplSEFL}
\end{equation}
Equations (\ref{eq:LaplSEFL}) are solved together with the Laplace
transforms of the boundary conditions ${\cal E}_{p}(0,t) = F(0,t)$,
where $F(0,t)$ is an incident probe pulse at the entry $x=0$ of the
PCM medium, and ${\cal E}_{c}(L,t) = 0$, so no incoming conjugate
pulse at the end of the medium. The result is
\begin{eqnarray}
\tilde{\cal E}_{c}(x,s) & = & h_{r}(x,is)\, \tilde{\cal F}_{*}(0,s)
\nonumber \\
\tilde{\cal E}_{p}(x,s) & = & h_{t}(x,is)\, \tilde{\cal F}(0,s),
\nonumber
\end{eqnarray}
with the reflection and transmission amplitudes
\begin{eqnarray}
h_{r} (x,\delta) & = & \frac{ |g| \sin(\beta (L-x))} {\frac{\delta}{c}
\mbox{\rm sin} (\beta L) + i \beta \mbox{\rm cos} (\beta L)}
\label{eq:conjrefl3} \vspace{0.3cm} \\
h_{t} (x,\delta) & =  & \frac{i \beta \cos (\beta (L-x)) + \frac{\delta}{c}
\sin (\beta (L-x))}
{\frac{\delta}{c}\mbox{\rm sin} (\beta L) + i \beta \mbox{\rm cos} 
(\beta L)}, 
\label{eq:probetrans3}
\end{eqnarray}
where 
\begin{equation}
\beta = \frac{1}{c} \sqrt{\delta^2 + |g c|^2}.
\label{eq:betavec}
\end{equation}

The reflected phase-conjugate pulse ${\cal E}_{c}(x,t)$ and
transmitted probe pulse ${\cal E}_{p}(x,t)$ at position $x$ in the
nonlinear medium and time $t$ are then obtained by using the inverse
Laplace transform. At $x=0$ and $x=L$, respectively, they are given by
\begin{eqnarray}
{\cal E}_{c}(0,t) & = & \frac{1}{2\pi i} \int_{\gamma - i\infty}^{\gamma + 
i\infty} ds\, r_{cp} (is) \, \tilde{{\cal E}}_{p^{*}}(0,s)\, e^{st} 
\label{eq:fisherr} \\
{\cal E}_{p}(L,t) & = & \frac{1}{2\pi i} \int_{\gamma - i\infty}^{\gamma + 
i\infty} ds\, t_{pp} (is) \, \tilde{{\cal E}}_{p}(0,s)\, e^{st},
\label{eq:fishert}
\end{eqnarray}
where $r_{cp} = h_{r}(x=0)$ and $t_{pp} = h_{t}(x=L)$.  The choice of
contour $\gamma$ is in agreement with causality, which means in this
case that it is to the right of all the singularities of $r_{cp}(is)$
(and $t_{pp}(is)$).  The singularities in the right-half s-plane give
rise to exponential growth of ${\cal E}_{c}(x,t)$ and ${\cal
E}_{p}(x,t)$. The integrals (\ref{eq:fisherr}) and (\ref{eq:fishert})
can be evaluated by taking the contribution of these poles separately
and rewriting the remaining integral as a Fourier integral.

They can also be evaluated in another way, which is especially
insightful when regarding non-analytic, chopped probe pulses. To that
end we rewrite (\ref{eq:fisherr}) and (\ref{eq:fishert}) as
\begin{eqnarray}
{\cal E}_{c}(0,t) & = & \int d t^{'}\, {\cal E}_{p}^{*}(0, t^{'}) 
H_{c}(t, t^{'}) \\
{\cal E}_{p}(L,t) & = & \int d t^{'}\, {\cal E}_{p}(0, t^{'}) H_{p}(t, t^{'}),
\label{eq:probeana}
\end{eqnarray}
with
\begin{eqnarray}
H_{c}(t, t^{'}) & = & \frac{1}{2\pi i}\int_{\gamma - i\infty}^{\gamma + i\infty}
ds\, r_{cp}(is)\, e^{-s(t^{'} - t)} \\
H_{p}(t, t^{'}) & = & \frac{1}{2\pi i}\int_{\gamma - i\infty}^{\gamma + i\infty}
ds\, t_{pp}(is)\, e^{-s(t^{'} - t)}
\label{eq:Hana}
\end{eqnarray}
In order to evaluate the integral in, e.g., $H_{p}(t, t^{'})$,
$t_{pp}$ is first rewritten as the series
\begin{equation}
t_{pp}(is) = 2\eta \sum_{n=0}^{\infty} \frac{a^{2n}}{(s+\eta)^{2n+1}}\,
e^{-(n + \frac{1}{2})\eta \tau}
\label{eq:series}
\end{equation}
with $a\equiv \kappa_{0} c$, $\eta \equiv \sqrt{s^2 - a^2}$ and $\tau
= 2L/c$, the roundtrip time of the PCM. One can then easily prove that
(\ref{eq:series}) is uniformly convergent, which allows for term by
term integration of (\ref{eq:Hana}). We use the substitution
\begin{equation}
s = \frac{i a}{2} \left( \frac{u}{A_{n}} - \frac{A_{n}}{u} \right)
\end{equation}
with
\begin{equation}
A_{n} \equiv \left( \frac{t - t^{'} - (n+\frac{1}{2}) \tau }
{t - t^{'} + (n+\frac{1}{2}) \tau } \right)^{1/2},
\end{equation}
in which $t^{'} + (n+\frac{1}{2})\tau$ is the retardation time after
$(n+\frac{1}{2})$ round trips. We then arrive at the following form of
the conjugate reflected pulse
\begin{eqnarray}
{\cal E}_{c}(L,t) & = & \int_{-\infty}^{t} d t^{'}\, 
{\cal E}_{p}^{*}(0,t^{'})\, G_{0}(t, t^{'}) + \nonumber \\
& & - 2 \sum_{n=1}^{\infty} \int_{-\infty}^{t - n \tau}
d t^{'} {\cal E}_{p}^{*}(0,t^{'})\, G_{n}(t, t^{'}),
\label{eq:Lfunctionconj}
\end{eqnarray}
with
\begin{equation}
\begin{array}{rl}
G_{n}(t, t^{'}) & = \frac{-i\, \kappa_{0} c}{4}  
\left(  
B_{n}^{n-1} I_{2n -2} \left[ |g| c\, \sqrt{(t - t^{'})^2  
- n^2 \tau^2}\, \right]  \right.
\vspace{0.2cm} \\
&  - 2 B_{n}^{n} I_{2n}\left[ |g| c\, \sqrt{(t - t^{'})^2 - 
n^2 \tau^2}\, \right]
\vspace{0.2cm} \\ 
& \left. + B_{n}^{n+1} I_{2n + 2} \left[ |g| c\,
\sqrt{(t - t^{'})^2 - n^2 \tau^2}\, \right] \ \ \right),
\end{array}
\end{equation}
and
\begin{equation}
B_{n} \equiv \left( \frac{t - t^{'} - n\, \tau }
{t - t^{'} + n\, \tau } \right).
\end{equation}


\begin{references}
\bibitem{brillouin60} L. Brillouin, {\em Wave Propagation and Group
                 Velocity}, (Academic Press, New York, 1960).
\bibitem{chu82} C.G.B. Garrett and D.E. McCumber, Phys. Rev. A {\bf
                1}, 305 (1970); S. Chu and S. Wong,
                Phys. Rev. Lett. {\bf 48}, 738 (1982).
\bibitem{buettiker82} M. B\"{u}ttiker and R. Landauer,
		Phys. Rev. Lett. {\bf 49}, 1739 (1982); Th. Martin and
		R. Landauer, Phys. Rev. A {\bf 45}, 2611 (1992).
\bibitem{steinberg92} A. M. Steinberg, P.G. Kwiat, and R.Y. Chiao,
 		Phys. Rev. Lett.  {\bf 68}, 2421 (1992).
\bibitem{ranfagni93} A. Ranfagni, D. Mugnai, and A. Agresti,
                  Phys. Lett. A {\bf 175}, 334 (1993).
\bibitem{japha96} Y. Japha and G. Kurizki, Phys. Rev. A {\bf 53}, 586
               (1996).
\bibitem{picholle91} E. Picholle, C. Montes, C. Leycuras, O. Legrand,
                  and J. Botineau, Phys. Rev. Lett. {\bf 66}, 1454
                  (1991).
\bibitem{icsevgi69} A. Icsevgi and W.E. Lamb, Phys. Rev. {\bf 185},
                517 (1969).
\bibitem{bolda96} E. L. Bolda, Phys. Rev. A {\bf 54}, 3514 (1996);
	        E. L. Bolda, J.C. Garrison, and R.Y. Chiao, {\it ibid}
	        {\bf 49}, 2938 (1994).
\bibitem{artoni98} M. Artoni and R. Loudon, Phys. Rev. A {\bf 57}, 622
               (1998).
\bibitem{aharonov69} Y. Aharonov, A. Komar, and L. Susskind,
		Phys. Rev.  {\bf 182}, 1400 (1969).
\bibitem{chiao96} R.Y. Chiao, A.E. Kozhekin, and G. Kurizki,
		Phys. Rev. Lett.  {\bf 77}, 1254 (1996).
\bibitem{blaauboer} G.~Kurizki, A.~Kozhekin and A.G.~Kofman,
        Europhysics Letters (in press); M.~Blaauboer, A.E.~Kozhekin,
        A.G.~Kofman, G.~Kurizki, D.~Lenstra, and A.~Lodder,
        Opt. Comm. {\bf 148}, 295 (1998); M.~Blaauboer, A.G.~Kofman,
        A.E.~Kozhekin, G.~Kurizki, D.~Lenstra, and A.~Lodder,
        Phys. Rev. A. {\bf 57} (in press).
\bibitem{barnett} S.M.Barnett and P.L.Knight, J.Opt.Soc.Am.B {\bf 2},
   467 (1985); C.M.Caves and B.L.Schumaker, Phys.Rev.A {\bf 31}, 3068,
   3093 (1985).
\bibitem{wang} B.S.~Wang and J.L.~Birman, Phys.Rev.B {\bf 42}, 9609 (1990)
          show numerically analogous features in the anomalous
          dispersion of phonoritons, namely, exciton-polaritons
          modified by the exciton-phonon interaction.
\bibitem{kirzhnits} A.Yu.~Andreev and D.A.~Kirzhnits Sov.Phys.Uspekhi {\bf
    39}, 1071, (1996); D.A.~Kirzhnits [Sov.Phys.Uspekhi] Usp.Fiz.Nauk
    {\bf 90}, 129, (1966); P.Csonka, Nucl.Phys. B {\bf 21}, 436
    (1970); R.Fox, C.Kuper, S.Lipson, Proc.Roy.Soc. A {\bf 316}, 515
    (1970); Y.Aharonov et al., Phys.Rev. {\bf 182}, 1400 (1969).
\bibitem{caves} C.M.~Caves and P.~Drummond, Rev.Mod.Phys. {\bf 66}, 481
    (1994).
\bibitem{raymer} M.G.~Raymer and J.~Mostowski, Phys.Rev.A {\bf 24}, 1980,
     (1981); M. G. Raymer, J. Mostowski, and J. L. Carlsten,
     Phys.Rev.A {\bf 19}, 2304 (1979).
\bibitem{feinberg} G.~Feinberg, Phys.Rev. {\bf 159}, 1089 (1967);
     O.~Bilaniyk and E.~Sudarshan, Phys.Today {\bf 22}, 43 (1969);
     S.~Weinberg, Phys.Rev.Lett. {\bf 19}, 12260 (1967); R.~Folman and
     E.~Recami, Found.Phys.Lett. {\bf 8}, 127 (1995); E.~Recami,
     Rivista Nuovo Cim. {\bf 9}, issue 6 (1986).
\bibitem{fisher81} R.A. Fisher, B.R. Suydam, and B.J. Feldman,
        Phys. Rev A {\bf 23}, 3071 (1981).
\end{references}
\end{document}